\documentclass[prd, twocolumn, lengthcheck,superscriptaddress, showpacs, letterpaper, footinbib]{revtex4}
\usepackage{ifpdf}
\usepackage[bookmarks=false]{hyperref}
\pdfoutput=1
\usepackage[usenames,dvipsnames]{color}
\usepackage{graphicx}
\usepackage{graphics}  
  \usepackage{float}
\usepackage{amsmath}
\usepackage{acronym}
\usepackage{multirow}
\usepackage{xspace}
\usepackage{units}
\usepackage{dcolumn}
\usepackage{ulem}

\newcommand{\ilya}[1]{{\color{black} #1}}
\newcommand{\jon}[1]{{\color{black} #1}}
\newcommand{\carl}[1]{{\color{black}  #1}}
\newcommand{\jontext}[2]{#2}

\begin{document}
\title{Verifying the no-hair property of massive compact objects with intermediate-mass-ratio inspirals in advanced gravitational-wave detectors}
\author{Carl L. Rodriguez}
\affiliation{Center for Interdisciplinary Exploration and Research in Astrophysics (CIERA) \& Dept.~of Physics and Astronomy, Northwestern University, 2145 Sheridan Rd, Evanston, IL 60208, USA
}
\email[]{carlrodriguez2015@u.northwestern.edu}
\author{Ilya Mandel}
\affiliation{NSF Astronomy and Astrophysics Postdoctoral Fellow, MIT Kavli Institute, Cambridge, MA 02139} 
\affiliation{School of Physics and Astronomy, University of Birmingham, Edgbaston, Birmingham, B15 2TT}
\author{Jonathan R.~Gair}
\affiliation{Institute of Astronomy, University of Cambridge, Madingley Road, Cambridge, CB3 0HA, UK }
\date{\today}
\pacs{04.80.Nn, 04.25.Nx, 04.30.Db, 04.80.Cc }

\begin{abstract}
The detection of gravitational waves from the inspiral of a neutron star or stellar-mass black hole into an intermediate-mass black hole (IMBH) promises an entirely new look at strong-field gravitational physics.  Gravitational waves from these intermediate-mass-ratio inspirals (IMRIs), systems with mass ratios from $\sim$10:1 to $\sim$100:1, may be detectable at rates of up to a few tens per year by Advanced LIGO/Virgo and will encode a signature of the central body's spacetime.  Direct observation of the spacetime will allow us to use the ``no-hair'' theorem of general relativity to determine if the IMBH is a Kerr black hole (or some more exotic object, e.g. a boson star).  Using modified post-Newtonian (pN) waveforms, we explore the prospects for constraining the central body's mass-quadrupole moment in the advanced-detector era.  We use the Fisher information matrix to estimate the accuracy with which the parameters of the central body can be measured.  We find that for favorable mass and spin combinations, the quadrupole moment of a non-Kerr central body can be measured to within a $\sim15\%$ fractional error or better using 3.5 pN order waveforms; on the other hand, we find the accuracy decreases to $\sim 100\%$ fractional error using 2 pN waveforms, except for a narrow band of values of the best-fit non-Kerr quadrupole moment.
\end{abstract}

\maketitle
\section{Introduction}
Within the coming decade, the field of gravitational wave astrophysics will become an effective tool for exploring the universe.  As gravitational waves are emitted from any system with a time dependent quadrupole moment (such as the inspiral and merger of two compact objects), this new field will provide us with an insight into the universe independent of conventional electromagnetic observations.  For the first time, we will be able to directly detect the coalescence of compact objects, such as neutron stars (NS) and black holes (BH). Of particular interest is the information about strong-field gravitational physics that these gravitational waves contain.  The analysis of the final moments of binary inspiral will provide us with the first direct tests of general relativity in the strong-field regime \cite{CutlerThorneReview}.

With the LIGO and Virgo gravitational wave observatories scheduled to reach their advanced sensitives \cite{AdvLIGO, AdvVirgo} around 2015, a large effort is underway to develop techniques capable of estimating the parameters of a coalescing compact binary composed of neutron stars and/or black holes.  Most of this work has been geared towards the detection of inspirals \jontext{of}{in which the two components are of} approximately equal mass, with the total mass of the system less than $\approx 35 M_{\odot}$ \cite{LSC4,LSC5}; however, more recent searches in the high-mass regime have \jontext{probed}{looked} for inspirals with total mass \jontext{from}{in the range} $25M_{\odot} \leq M \leq 100M_{\odot}$ and component masses \jontext{from}{in the range}  $1M_{\odot} \leq m_1\text{, }m_2 \leq 99M_{\odot}$, searching for more extreme mass ratios \cite{HighMass}.  These systems emit gravitational waves in the frequency range best suited for detection by Advanced LIGO, the peak sensitivity of which will lie at $\sim 100$ Hz.

Both observational evidence and theoretical models suggest the existence of intermediate-mass black holes (IMBHs) with masses ranging from under a hundred to ten thousand solar masses (see \cite{MillerColbert:2004} for a review).  While the recently discovered ultra-luminous X-ray source in ESO 243-49 may turn out to be the first confident IMBH detection \cite{ObserveIMBH}, many questions about IMBH formation mechanisms and their prevalence in dense stellar environments remain unanswered.  Gravitational waves emitted during the intermediate-mass-ratio inspirals (IMRIs) \jontext{}{of stellar-mass compact objects into IMBHs} will be in the observable frequency band of Advanced LIGO \cite{BrownIMRI} and dynamics studies suggest possible IMRI detection rates of up to a few tens per year \cite{IlyaRates}. 

The gravitational waveforms generated by these inspirals will encode information about the higher-order multipole moments of the central body's spacetime.  Since the ``no-hair'' theorem predicts that all the higher-order multipole moments of a Kerr black hole can be expressed as a function of its mass and spin, a simultaneous measurement of the mass, spin, and mass-quadrupole moment would serve as a test of the null hypothesis that massive compact objects are Kerr black holes and inspirals are governed by general relativity.

In this paper, we describe the potential for constraining the mass-quadrupole moment of the central body in an IMRI.  This idea was first rigorously developed by Ryan, who showed that the inspiral of a nearly circular, nearly equatorial extreme mass ratio binary encodes information about \textit{all} the higher order mass and spin moments of the spacetime \cite{RyanMultipole}.  While Ryan studied the detectability of all higher order multipole moments \cite{RyanDetection}, more recent work has focused on isolating the mass-quadrupole moment (IMRIs in LIGO by Brown et al. \cite{BrownIMRI}, and extreme-mass-ratio inspirals in LISA by Barack and Culter \cite{BCEMRI}).  Whereas previous work focused on \jontext{}{deviations about} the null case of a Kerr spacetime, we offer the first exploration of the quadrupole moment parameter space, looking at non-Kerr objects and comparing their detectability to that of the null case.   For simplicity, we restrict ourselves to the \carl{quasi-circular, single-spin} case where the orbital and IMBH angular momenta are aligned.

Using the Fisher matrix formalism, we \jontext{measure the upper limit uncertainties on}{compute the precision of} parameter estimation \jontext{given the}{that will be possible using} Advanced LIGO \jontext{design noise}{operating at its design} sensitivity.  By including an additional term in the waveform representing deviations from the theoretically predicted quadrupole moment, we \jontext{will}{find that we should} be able to constrain any deviation of the quadrupole moment with a fractional error of $\sim 1-10$ when the best-fit parameters are those of a Kerr central body, and to $\sim 15\%-300\%$ or better for certain non-Kerr central bodies, depending on our choice of pN order and central-body mass and spin.  In some cases, this will make it possible to discern between a Kerr black hole and some more exotic, horizon-less object, such as a boson star \cite{RyanBoson}.  If the object \textit{is} a Kerr black hole, we will be able to \jontext{determine how well}{demonstrate that} it matches the predictions of general relativity \jontext{}{to the same precision}.  

In section~\ref{IMRIrate}, we review the current theoretical understanding of IMBH formation and IMRIs, with a particular \jontext{eye to}{focus on} detection rates in Advanced LIGO.    \jontext{Section~\ref{IMRIwave} we cover}{In section~\ref{IMRIwave}  we describe} the frequency domain waveforms and their numerical implementation in the Fisher information matrix (FIM).  Section~\ref{results}  reports the results of two cases: the null case (a Kerr IMBH), and the anomalous case (in which the best-fit quadrupole moment deviates from the no-hair Kerr IMBH).  We also discuss technical issues arising from the FIM, as well as our use of the post-Newtonian waveforms for IMRI systems.  Unless otherwise stated we use geometrized units, with $G=c=1$.

\section{Gravitational Waves from IMRIs}
\label{IMRIrate}
\subsection{Rate Estimates}

Numerical and observational studies of globular clusters (including ultra-luminous X-ray sources) have suggested the existence of IMBHs with masses from $10^2 M_{\odot}$ to $10^4 M_{\odot}$.  Unlike stellar mass black holes, the formation mechanisms for these objects are not well understood.  Currently, one of the most promising formation mechanisms is the runaway merger scenario.  By \jontext{repeatedly combining}{successively colliding} massive stars on a timescale much faster than that of stellar evolution ($\lesssim 3$ Myr), one can \jontext{essentially}{} grow a massive star (800-3,000$M_{\odot}$), which then collapses to an IMBH \cite{PortegiesIMBH}.  However, more recent simulations have suggested that stellar winds would prevent this mechanism in any but the most metal-poor environments~\cite{Glebbeek09}.  It is also possible that IMBHs could form through a series of mergers with other compact objects in \jontext{the}{a} dense sub-cluster in the center of a globular cluster.  The issue with this process is that these collisions would generate objects with high \jontext{product}{kick} velocities, most likely leading them to be ejected from the cluster \cite{OlearyIMBH,BanerjeeBH}; however, with a sufficiently heavy seed black hole in the cluster center, possibly formed by the above mechanism, these IMBHs could be retained \cite{IlyaRates}.  

There are other possible, albeit less explored, mechanisms for IMBH formation.  It is possible that 
early, massive, low-metallicity population III stars could collapse directly into an IMBH because of the greatly reduced mass loss via stellar winds \cite{VolonteriSMBH}.  It is also possible that gas accretion could \jontext{yield}{lead to} direct IMBH formation early in a globular cluster's history.  \jontext{By surrounding standard}{A} solar mass black hole in a gas rich environment\jontext{, they}{} would eventually accrete enough \jontext{}{mass} to enter the intermediate mass regime \cite{PortegiesGas,VesperiniGas}.  Despite these possibilities, the exact processes leading to IMBH formation are unclear.  Direct detection of gravitational waves from a merger involving an IMBH could provide the first irrefutable proof of their existence, giving us a greater insight into the nature of these mysterious objects \cite{S4Ringdown}.  

\jontext{With this, Mandel et al.~produced possible IMRI rates detectable in the advanced detector era, given the following assumptions.  }{If IMBHs do indeed exist at a sufficiently high density, their mergers with stellar mass compact objects will lead to IMRI signals in Advanced LIGO. The rate of detectable IMRI signals was estimated in~\cite{IlyaRates} under the following set of assumptions.} First, that the Advanced LIGO network will have a lower frequency cutoff of 10Hz, with a minimum signal-to-noise ratio of 8 \jontext{}{required} for a coherent detection.  Secondly, that the number density of globular clusters with a sufficiently high central density that contain IMBHs is $\sim 0.3(g/0.1)\text{Mpc}^{-3}$, with a high uncertainty of the fraction $g$, although simulations suggest that $g\sim 10\%$ for IMBHs up to $300M_{\odot}$ may be plausible.  Finally, that the rate of IMRIs per globular cluster is driven primarily by three-body hardening of a compact-object--IMBH pair.  This scenario yields an estimated rate per cluster of $\alpha\sim 10^{-9}(g/0.1)\text{ Mpc}^{-3}\text{ yr}^{-1}$ for both NS-IMBH and 
%$\alpha\sim 1.5\times10^{-9}(f/0.1)\text{ MPC}^{-3}\text{ yr}^{-1}$ for 
BH-IMBH systems.  \jontext{Combining}{Under} these three assumptions, \jontext{Mandel et al. concluded that}{} Advanced LIGO could reasonably observe one NS-IMBH event every 3 years, and up to 10 BH-IMBH inspirals per year~\cite{IlyaRates}.  With sufficient improvements to the low-frequency sensitivity, that number could improve to 1 and 30 events per year for NS and BH inspirals, respectively.

\subsection{Why IMRIs?}

As stated above, most searches for gravitational waves from compact binary coalescence have focused on the equal mass regime.   \jontext{Currently the most sensitive part of the}{The current} proposed Advanced LIGO noise curve \jontext{lies from approximately}{has greatest sensitivity in the} 100Hz to 500Hz \jontext{}{range}, with a peak sensitivity \jontext{of}{at} $\sim$250Hz.  In the case of a NS-NS binary, where the plunge and merger of the bodies occurs at $\sim$1600Hz, only the intermediate portion of the inspiral falls into this most sensitive region.  However, for strong-field tests, we are interested in the effects that come from the final moments of inspiral when the companion object is very close to the central body.  What we want is a signal that terminates near the maximum sensitivity point of our detector, ensuring we are \jontext{most}{as} sensitive \jontext{}{as possible} to the higher-order post-Newtonian effects.  Since our upper frequency cutoff for binary inspirals is inversely proportional to the total mass (in the non-spinning case), we expect that an IMRI which terminates just slightly above 250Hz (corresponding to a total mass in the $10-60M_{\odot}$ range) will be ideal for our purposes.

At the same time, a more extreme mass ratio means that the compact object ``test particle'' spends more orbital cycles close to the horizon of the IMBH, gathering more information about the strong-field of the central body.  Furthermore, asymmetric mass ratios may allow us to ignore complicating contributions from the companion, such as companion spin.  As such, we expect our ideal systems to be IMRIs with a moderate total mass and high mass ratio.  This will ensure that the inspiral produces a large number of strong-field cycles in the highest sensitivity section of the detector bandwidth.

\section{Fisher Matrix Implementation}
\label{IMRIwave}
Here we present the formalism for determining the expected precision of parameter estimation using the Fisher information matrix.  This technique has been regularly applied to gravitational wave parameter estimation (e.g. \cite{FinnDetection,CutlerFlanagan}).  For our purposes, we follow the setup detailed in Poisson and Will \cite{PoissonWill}.

\subsection{Modified Waveform}

We use a frequency domain waveform accurate up to $3.5$ post-Newtonian (pN) order (sometimes called the \textit{TaylorF2} approximant \cite{BuonannoWaveform}).  Using the stationary phase approximation, the gravitational wave amplitude is given by

\begin{equation}
\tilde{h}(f) = A f^{-7/6}e^{i \psi(f)},
\label{amplitude}
\end{equation}

\noindent where $A \propto \mathcal{M}_c^{5/6}\Theta(\text{angle})/D$, $D$ is the luminosity distance of the binary, and $\psi(f)$ is the pN phase.  $\Theta(\text{angle})$ is a function of the orbital orientation with respect to the detector network in terms of the sky position, orbital inclination, and the wave polarization.  In addition to the total mass, $M\equiv m_1+m_2$, it is convenient to work with the \textit{mass ratio} and \textit{chirp mass}, defined by

\begin{equation}
\eta\equiv m_1m_2/M^2~~~~\text{and}~~~~\mathcal{M}_c = \eta^{3/5}M,
\end{equation}

\noindent respectively.  Then, in terms of the Newtonian orbital velocity $v=(\pi M f)^{1/3}$, the 3.5pN phase is 

\begin{equation}
\psi(f) = 2 \pi f t_c - \phi_c + \frac{\pi}{4} + \frac{3}{128 \eta}v^{-5}\sum^{7}_{k=0}\alpha_{k}v^k
\label{phase}
\end{equation}

\noindent with coefficients

\begin{align}
\alpha_0 &= 1\\
\alpha_1 &= 0 \nonumber\\
\alpha_2 &= \frac{20}{9}\left(\frac{743}{336}+\frac{11}{4}\eta\right)\nonumber\\
\alpha_3 &= -4(4\pi - \beta)\nonumber\\
\alpha_4 &= 10\left(\frac{3058673}{1016064}+\frac{5249}{1008}\eta+\frac{617}{144}\eta^2  - \sigma_{qm} \right)\nonumber\\
\alpha_5 &= \pi\left(\frac{38645}{756}-\frac{65}{9}\eta\right)\left(1+3\log\left(\frac{v}{v_{\text{ISCO}}}\right)\right)\nonumber\\
\alpha_6 &= \frac{11583231236531}{4694215680} - \frac{640}{3}\pi^2 - \frac{6848}{21}\gamma_{\text{euler}} \nonumber\\
         &- \frac{6848}{21} \log\left(4 \frac{v}{v_{\text{ISCO}}}\right) -\left(\frac{15737765635}{3048192}-\frac{2255}{12}\pi^2\right)\eta\nonumber\\
           &+ \frac{76055}{1728}\eta^2 - \frac{127825}{1296}\eta^3\nonumber\\
\alpha_7 &= \pi\left(\frac{77096675}{254016}+\frac{1014115}{3024}\eta - \frac{36865}{368}\eta^2\right)\nonumber
\end{align}

\noindent where $\gamma_{\text{euler}}\approx 0.577$ and $v_{\text{ISCO}}$ is the orbital velocity at the cutoff frequency.  \carl{We restrict ourselves to quasi-circular waveforms as a simplifying assumption.  While}\ilya{, in principle, eccentricity could significantly impact the waveforms \cite{DuncanEccentricity},}
%we could incorporate eccentricity following the example in \cite{DuncanEccentricity} , 
\carl{we expect IMRI systems to have negligible eccentricities when formed via three-body hardening \cite{IlyaRates}.}  The terms $t_c$ and $\phi_c$ in equation (\ref{phase}) are constants of integration, referring to the chirp time and chirp phase, respectively.  Although uninteresting physically, they must be included in any parameter estimation study of the waveform phase.  The terms $\beta$ and $\sigma_{qm}$ represent the leading-order spin-orbit and spin-quadrupole interactions of the two masses.  These contributions take the form

\begin{align}
\beta &= \frac{1}{12}\sum^{2}_{i=1}\big(113(m_i/M)^2 + 75\eta\big)\mathbf{\hat{L}\cdot\chi_i}\\
\sigma_{qm} &= -\frac{5}{2}\sum^{2}_{i=1}\frac{Q_i}{m_i M^2}\left(3(\mathbf{\hat{L}\cdot\hat{\chi_{i}}})^2 - 1\right)
%\sigma_{\text{self}}&=\frac{1}{96 M^2}\sum^{2}_{i=1}(\chi_i m_i)^2(6 + \sin^2 (\mathbf{\hat{L}\cdot\hat{\chi_{i}}}))\\
\end{align}

\noindent where $\mathbf{\chi_i}$ is the dimensionless spin of the $i^{th}$ companion, and 

\begin{equation}
Q_{i} = (-\chi_i^2 + Q_{\text{anom}}) m_i^3
\label{quadrupole}
\end{equation}

\noindent is the relativistic mass-quadrupole moment (since we are only interested in the quadrupole moment of a single object, we ignore the indices and only include terms from the central body).  We have introduced the dimensionless parameter $Q_{\text{anom}}$ to characterise a non-Kerr quadrupole moment.  \carl{This deviation is similar to the one introduced for supermassive black holes by Barak and Cutler in} \ilya{their analysis of LISA extreme-mass-ratio inspirals \cite{BCEMRI}.}   In general, there are three additional terms to $\sigma$, representing the spin-spin, self-spin, and dipole-dipole interactions \cite{MikocziSelf}.  However, since we have restricted ourselves to IMRIs where the central body spin is dominant, we ignore the spin-spin and dipole coupling effects.  Additionally, the self-spin correction is proportional to $\sigma_{qm}$ when $Q_{\text{anom}} $ is 0, but its magnitude is $\sigma_{\text{self-spin}}/\sigma_{qm} = 1.25\%$, and in practice we found it had no noticeable effect on parameter estimation. 
% \ilya{Interesting, I've never heard of spin-self interaction, and Poisson doesn't seem to include it... Jon, do you know anything about this?}\carl{I found it in \cite{MikocziSelf} while looking for higher order spin terms; it was calculated significantly after Poisson's stuff, so far as I can tell.}\jon{I was dimly aware of it, but assumed it would enter at higher PN order. Surprising that it can contribute as much as the others. Does this come from the Papapetrou equation somehow?} \carl{The paper isn't very specific, but a quick look through the references says it comes from the 2pN correction to Lense-Thirring precession.  It's not physically clear to me why the effect persists with only one spinning object}

In this paper, we \carl{make several simplifying assumptions in the construction of our template family.  In addition to circular orbits, we have explicitly assumed
%\begin{itemize}
%\item 
(i) that the spin of the companion object can be ignored; and
%\item 
(ii) that the IMBH spin is aligned with the orbital angular momentum.}
%\end{itemize}
 %explicitly set the spin of the companion object to zero, and examine only the cases where the central body spin is aligned with the orbital angular momentum. 
%\noindent 
%These assumptions reduce us to a 6-dimensional subspace of the full 12-dimensional parameter space (i.e.\ $m_1$, $m_2$, $\vec{\chi}_1$, $\vec{\chi}_2$, $Q_{\text{anom}}$, $e$, $\phi_c$, and $t_c$).} 
%The full space has more dimensions -- e.g., in addition to eccentricity, there's the argument of the periapsis.  How about just 
\ilya{These assumptions significantly reduce the dimensionality of our parameter space, which, in principle, could have up to 17 parameters: the two component masses, the two spin vectors, the eccentricity and argument of periapsis, the location of the binary on the sky, the direction of its orbital angular momentum, and the phase and time of coalescence -- as well as the additional $Q_{\text{anom}}$.}  
%Of course, there is no physical reason to expect this to be the case in nature: 
\carl{The spin distribution of IMBHs is almost entirely unconstrained at present, and for dynamically formed binaries 
%it is perfectly conceivable that 
both the IMBH and companion object could have spins misaligned with the orbital angular momentum,}
 %While our simplification is purely for convenience,
\ilya{although a sufficiently extreme mass ratio makes the spin of the companion increasingly unimportant \cite{MandelGairDetection}.  The possible impact of including spin-spin and spin-orbit precession on the accuracy of parameter estimation is difficult to estimate.  On the one hand, increasing the dimensionality of the parameter space tends to decrease the accuracy of measuring individual parameters because of inter-parameter correlations.  On the other hand, the waveform modulation from precession contains additional information that can break degeneracies, potentially leading to improved measurement accuracy \cite{Sluys08}.}
%practice the presence of spin-spin and spin-orbit precession improves parameter estimation due to the additional information encoded in the waveforms \cite{Sluys08}, suggesting that at worst our assumptions are conservative.  [THAT MIGHT BE A STRETCH] }

Our waveforms range from the lower frequency Advanced LIGO cutoff of 10Hz to the innermost stable circular orbit (ISCO) frequency of the system, which we define by

\begin{equation}
\pi f_{\text{ISCO}} = \frac{M^{1/2}}{r_{\text{ISCO}}^{3/2}+\chi_1 M^{3/2}}
\label{ISCO}
\end{equation}

\noindent where $r_{\text{ISCO}}$ is the separation of the two masses at ISCO \cite{BPT}, given by

\begin{align}
r_{\text{ISCO}}/M &= 3 + Z_2 \mp \sqrt{(3-Z_1)(3+Z_1 + 2 Z_2)}\label{rISCO}\\
Z_1 &\equiv 1 + (1 - \chi_1^2)^{1/3}\nonumber\\
&~~~~~\times \left[(1+\chi_1)^{1/3} + (1 - \chi_1)^{1/3} \right]\nonumber\\
Z_2 &\equiv\sqrt{3\chi_1^2 + Z_1^2}\nonumber
\end{align}  

\noindent For prograde orbits, equation (\ref{rISCO}) varies from $r_{\text{ISCO}}=6M$ in the non-spinning case to $r_{\text{ISCO}}=M$ in the fully spinning case. For a non-spinning 1$M_{\odot}$ system, this corresponds to an $f_{\text{ISCO}}$ of ~4.4KHz. \jontext{}{The most sensitive part of the Advanced LIGO noise curve, at $250$Hz, is the ISCO frequency of a $\sim17.5M_\odot$ non-spinning black hole and it is the ISCO frequency of a $\sim58M_\odot$ black hole with spin $\chi=0.9$.}

We use the noise power spectral density, $S_{n}(f)$, provided by the LIGO Scientific Collaboration, representing the best estimate for a high-power, zero-detuning configuration of Advanced LIGO.  See \footnote{Both the noise curve and technical reports describing it can be found under \href{''https://dcc.ligo.org/cgi-bin/DocDB/ShowDocument?docid=2974''}{LIGO Document T0900288-v3}} for a more complete description of the sensitivity curves.

Finally, we note that the signal-to-noise ratio (SNR) of a gravitational wave is defined as 

\begin{equation}
\rho \equiv \frac{4}{\sigma} \int^{\infty}_{0}\frac{| \tilde{s}(f)\tilde{h}^{*}(f)|}{S_{n}(f)}df
\label{formalSNR}
\end{equation}

\noindent where $\rho$ is the SNR and $\tilde s(f)$ and $\tilde{h}^{*}(f)$ are the frequency domain signal and template, respectively.  The normalization $\sigma$ is given by

\begin{equation}
\sigma^2 = 4\int^{\infty}_{0}\frac{| \tilde{h}(f)|^2}{S_n(f)}df
\label{SNRnorm}
\end{equation}

\noindent Since we lack a true signal with which to use equation (\ref{formalSNR}), we approximate the SNR using $\sigma$ from equation (\ref{SNRnorm}) assuming Gaussian noise, for a network of detectors as 

\begin{equation}
\rho = \sqrt{\sum_i \sigma_i^2}
\label{snr}
\end{equation}

\noindent where the index $i$ refers to the signal and noise spectrum of the $i^{\text{th}}$ detector.  We choose the distance of each source to yield a network SNR of $\rho=20$.  We consider the complete advanced detector network consisting of the two LIGO sites (in Hanford, WA and Livingston, LA) and the Virgo site (in Pisa), although for simplicity we use the Advanced LIGO sensitivity for all three detectors.

\subsection{Analytic Setup}

Let us assume detection of a gravitational wave by a network of detectors, given by

\begin{equation}
s_i(t) = h_i(t;\boldsymbol{\theta}) + n_i(t) ,
\label{signal}
\end{equation}

\noindent where $s_i$ is the \jontext{detector}{} output \jontext{of the $i$'th detector in the network}, a combination of the noise $n_i$, which we assume to be  stationary and Gaussian, and $h_i$, the \jontext{pure}{true} waveform with \jontext{input}{} parameters $\boldsymbol{\theta}$.  The posterior probability of such a signal being detected with the given parameters is

\begin{equation}
p(\boldsymbol{\theta}|s) \propto p(\boldsymbol{\theta})\exp\left[-\frac{1}{2}\left< h(\boldsymbol{\theta})-s ~\vline~ h(\boldsymbol{\theta})-s \right> \right] ,
\label{basicprob}
\end{equation}

\noindent where $p(\boldsymbol{\theta})$ is the prior probability, chosen from our physical knowledge of the parameter space.  The brackets, $\left< ~|~ \right>$, refer to the noise-weighted inner product of two signals in the frequency domain, defined as

\begin{equation}
\left<a~\vline~ b\right> \equiv 4 \Re \int^{f_{\text{isco}}}_{10}\frac{\tilde{a}(f)\tilde{b}^{*}(f)}{S_{n}(f)}df ,
\label{overlap}
\end{equation}

\noindent where $S_n(f)$ is the power spectral density \jontext{}{and $\Re$ denotes the real part}.  It can be shown \cite{FinnDetection}  that equation (\ref{basicprob}) can be expanded to leading order in SNR as

\begin{equation}
p(\boldsymbol{\theta}|s) \propto p(\boldsymbol{\theta})\exp\left[-\frac{1}{2}\Gamma_{ab} \Delta\theta^a \Delta\theta^b\right] ,
\label{fisherprob}
\end{equation}

\noindent where $\Delta\theta^a$ is the difference between the $a^{th}$ parameter and its maximum likelihood estimate, and $\Gamma$ is the Fisher information matrix, defined by

\begin{equation}
\Gamma_{ab} \equiv \left<\frac{\partial h}{\partial\theta^a}\Big|\frac{\partial h}{\partial\theta^b}\right> .
\label{fisher}
\end{equation}  

\noindent  Note that equation (\ref{fisher}) is technically evaluated at the best-fit parameters of our template, not the ``true'' values of the waveform, and it does not provide information about the \jontext{uncertainty}{difference} between the real and best-fit values.  

Having cajoled equation (\ref{fisherprob}) into the form of a Gaussian, the variance-covariance matrix, $\Sigma$, is simply the inverse of the Fisher matrix, i.e. $\Sigma^{ab}=(\Gamma^{-1})^{ab}$.  With this formalism in hand, we can write the standard deviations and correlations between parameters as 

\begin{align}
\sigma_a &= \sqrt{\Sigma^{aa}} \label{stdev} \\
c_{ab} &= \frac{\Sigma^{ab}}{\sqrt{\Sigma^{aa}\Sigma^{bb}}}\label{correlation}
\end{align}

\noindent Taken together, equations (\ref{stdev}) and (\ref{correlation}) provide a theoretical lower limit on  the covariance of the posterior probability distribution of the parameters, given a noise realization $S_{n}$.

While a \jontext{highly effective first pass}{useful first approximation to parameter estimation accuracies}, the Fisher information matrix does suffer from several issues.  The expansion \jontext{which yielded}{leading to} equation (\ref{fisher}) requires a sufficiently high SNR in order to remain valid.  Unfortunately, the requisite SNR is heavily dependent on one's choice of waveform and region of parameter space.  We choose $\rho = 20$ as a fiducial value, \jontext{}{in the hope that this is} sufficiently loud to avoid \jontext{most}{a} break down of the linear signal approximation; however, by using the pN waveforms at such  \jontext{exotic}{extreme} mass ratios, we run the risk of overloading the linear approximation.  Since this is a function of where we explore the parameter space, there is no simple way to correct this.

Additionally, the Fisher matrix operates poorly \jontext{in areas}{} where the parameter correlations cause the determinant of $\Gamma$ to be near singular.  We discuss the generic numerical issues produced by this effect in the next section, and the issues specific to our systems of interest in section IV C.  In particular, see Table \ref{cctable} and associated text for an example of parameter correlations specific to our waveform model/mass regime.  For an excellent and thorough review of the pitfalls of the Fisher matrix in gravitational wave signal analysis, see Vallisneri \cite{Vallisneri}.

\subsection{Numerical Implementation}
We use a ninth order \carl{finite difference scheme}  to numerically compute the partial derivatives of the waveform with respect to the parameters.  The step size of the derivative must be carefully chosen: too large a step size would \jontext{miss analytic details of the waveform}{take the waveform differences outside the linear regime}, while too small a size would introduce numerical errors.  To that end, each parameter step size was independently chosen by minimizing the local error of the derivative.  By computing the waveform derivatives over a logarithmic grid of different step sizes (from ~$10^{-10}$ to $0.1$), and computing the unnormalized $\chi^2$ of each derivative with respect to its two closest grid neighbors, we can calculate the relative error of each step size, thereby selecting the step size with a minimum error.

Once the partial derivatives are computed, we form the Fisher matrix using equation (\ref{fisher}), and then calculate the covariance of the parameters by numerically inverting the matrix.  This step requires care, as the Fisher matrix is often near singular, and if the condition number of the matrix (the ratio of the largest to smallest eigenvalues) exceeds the decimal precision limit of the machine, it will introduce numerical errors larger than the inverse Fisher elements themselves.  For 32 bit machines, the precision limit is about 15 decimal places, so if the condition number of the matrix exceeds ~$10^{16}$, the results are numerically unstable.  This can arise in two ways.  If our choice of units is poor for a particular parameter, then the difference in magnitude between any two columns can exceed the machine limit.  This can be easily corrected by normalizing each matrix column to a set value.  On the other hand, if two parameters are nearly perfectly correlated (e.g. distance and orbital inclination), then that sub-matrix will be close to singular (with a determinant \jontext{well beyond standard }{considerably smaller than} floating point precision).

As this effect often arises from physical degeneracies within the waveforms, there is no easy numerical solution. Analytically, one can sometimes chose parameters such that the Fisher matrix is nearly diagonal, eliminating some of the degeneracies.  However, \jontext{this requires a cleverness of parameters that}{such parameter combinations are usually not known analytically and diagonalisation} cannot easily be implemented numerically, as any attempt to \jontext{numerically diagonalize the matrix}{do so} would require computation beyond standard floating point precision.  To be conservative, we choose to only invert matrices with condition numbers below $10^{15}$, ensuring our results are free of rounding errors.  The effectiveness of this cutoff was tested by introducing small perturbations into the input parameters and ensuring they corresponded to small perturbations in $\Sigma$.

In addition, a small sample of the results in this paper were checked against two other methods of analyzing parameter uncertainties.  To confirm the stability of numerical derivatives, we checked some results against a Fisher matrix with analytically computed waveform derivatives, c.f. equation (3.10) in \cite{PoissonWill}.  We also performed a grid-based search in the four-dimensional space of masses, spin, and mass-quadrupole moment, automatically maximizing over phase and time of coalescence, and compared uncertainty estimates predicted by the Fisher matrix against drops in overlaps between injected waveforms and templates evaluated at offset parameter values.

\begin{table*}[bthp]
\caption{$1$-$\sigma$ parameter uncertainties as computed from the Fisher matrix using the TaylorF2 waveform to 2pN and 3.5pN order.  In both cases the spin and quadrupole moment terms are included up to 2pN order.  We consider the spin-aligned inspiral case, with a companion object of $1.4M_{\odot}$ (a typical neutron star mass) and a central body of various masses and spins.  The times are reported in milliseconds, the angles in radians, and all other quantities are dimensionless.}
\begin{tabular}{lcccccccccccccc}
\hline\hline
Central Body & \vline & \multicolumn{6}{c}{2pN} & \vline & \multicolumn{6}{c}{3.5pN}\\
\hline
 & \vline & $~\Delta\mathcal{M}_c/\mathcal{M}_c~$ & $~\Delta\eta/\eta~$ & $~\Delta t_c~$ & $~~\Delta\phi_c~~$ & $~\Delta \chi_1~$ & $~\Delta Q_{\text{anom}}~$ & \vline & $~\Delta\mathcal{M}_c/\mathcal{M}_c~$ & $~\Delta\eta/\eta~$ & $~\Delta t_c~$ & $~~\Delta\phi_c~~$ & $~\Delta \chi_1~$ & $~\Delta Q_{\text{anom}}~$ \\

\hline
$\chi_1 = 0$\\
\hline
$10M_{\odot}$ & \vline & 0.369\% & 1.319 & 2.4 & 36.1 & 0.510 & 1.925 & \vline & 0.205\% & 1.808 & 217.1 & 357.6 & 0.078 & 13.165 \\
$25M_{\odot}$ & \vline & 1.410\% & 2.982 & 12.0 & 116.6 & 1.130 & 2.546 & \vline & 0.137\% & 1.708 & 71.6 & 806.4 & 0.333 & 12.006 \\
$50M_{\odot}$ & \vline & 4.067\% & 6.198 & 49.5 & 332.6 & 2.491 & 3.459 & \vline & 0.896\% & 2.588 & 361.8 & 2509.1 & 0.875 & 19.118 \\
$100M_{\odot}$ & \vline & 5.903\% & 7.051 & 160.5 & 616.8 & 2.977 & 3.146 & \vline & 4.680\% & 4.574 & 2232 & 9314.6 & 2.342 & 36.542 \\
$150M_{\odot}$ & \vline & 5.995\% & 7.799 & 529.4 & 1301.6 & 3.017 & 5.119 & \vline & \multicolumn{6}{c}{Unstable}\\
\hline 
\hline
$\chi_1 = 0.5$\\
\hline
$10M_{\odot}$ & \vline & 0.278\% & 0.929 & 1.2 & 22.0 & 0.035 & 1.455  & \vline & 0.080\% & 0.834 & 7.3 & 171.6 & 0.396 & 6.113 \\
$25M_{\odot}$ & \vline & 0.918\% & 1.770 & 4.7 & 57.7 & 0.126 & 1.766 & \vline & 0.128\% & 0.688 & 26.2 & 343.1 & 0.416 & 4.867 \\
$50M_{\odot}$ & \vline & 2.621\% & 3.566 & 17.7 & 154.2 & 0.405 & 2.643 & \vline & 0.621\% & 0.849 & 107.4 & 888.6 & 0.665 & 6.318 \\
$100M_{\odot}$ & \vline & 5.414\% & 5.499 & 59.7 & 338.9 & 0.839 & 3.226 & \vline & 2.848\% & 1.417 & 618.1 & 3212.5 & 1.471 & 11.464 \\
$150M_{\odot}$ & \vline & 5.923\% & 5.361 & 121.9 & 471.3 & 0.894 & 3.007 & \vline & 5.148\% & 1.625 & 1493.0 & 5779.8 & 1.950 & 13.779 \\
\hline
\hline
$\chi_1 = 0.9$\\
\hline
$10M_{\odot}$ & \vline & 0.205\% & 0.634 & 0.5 & 12.6 & 0.189 & 0.931 & \vline & 0.032\% & 0.355 & 2.7 & 94.0 & 0.04 & 2.591 \\
$25M_{\odot}$ & \vline & 0.556\% & 0.956 & 1.4 & 24.4 & 0.167 & 0.918 & \vline & 0.113\% & 0.226 & 7.5 & 151.4 & 0.231 & 1.615 \\
$50M_{\odot}$ & \vline & 1.370\% & 1.611 & 4.1 & 52.0 & 0.177 & 1.217 & \vline & 0.359\% & 0.213 & 23.2 & 317.3 & 0.277 & 1.610 \\
$100M_{\odot}$ & \vline & 3.532\% & 2.982 & 14.3 & 127.0 & 0.179 & 1.914 & \vline & 1.203\% & 0.265 & 98.3 & 918.3 & 0.457 & 2.170 \\
$150M_{\odot}$ & \vline & 5.101\% & 3.635 & 27.4 & 189.1 & 0.146 & 2.184 & \vline & 2.530\% & 0.335 & 256.9 & 1926.5 & 0.688 & 2.890 \\
\hline\hline

\end{tabular}
\label{nulltable}
\end{table*}

\section{Results}
\label{results}
\jontext{Throughout}{In} this section, we report the results for IMRI systems with a wide range of input parameters, testing two different waveform orders: one with all known terms up to second post-Newtonian order, and one with the known quadrupole terms to 2pN, and all known mass terms to 3.5pN order.  Although the 2pN waveform is expected to be less accurate, the quadrupole moment correction is only known up to 2pN order, making it unclear whether including some terms to higher orders (i.e., the mass terms) \jontext{is necessarily superior}{will necessarily yield more reliable results}.  Moreover, we already know that both waveforms match poorly with the ``true'' waveforms in the IMRI mass regime \cite{MandelGairDetection}.  Therefore, comparing the results from both pN orders allows us to form a qualitative understanding of the true parameter space while also exposing the dependence of quantitative results on our choice of waveform template. 

We focus on two cases: the null case in which the best-fit parameters correspond to a central body which is a Kerr black hole ($Q_{\text{anom}}=0$) and we wish to know how accurately we can constrain deviations away from this value, and the anomalous case, in which the best-fit parameters correspond to a central body with a non-Kerr mass-quadrupole moment which we are attempting to identify.  The latter case can introduce additional concerns about the validity of the post-Newtonian expansion of the phase, which we discuss in section IV C.  

\subsection{Null Detectability}

\begin{figure}[htb]
 \centering
 \includegraphics[scale=0.58]{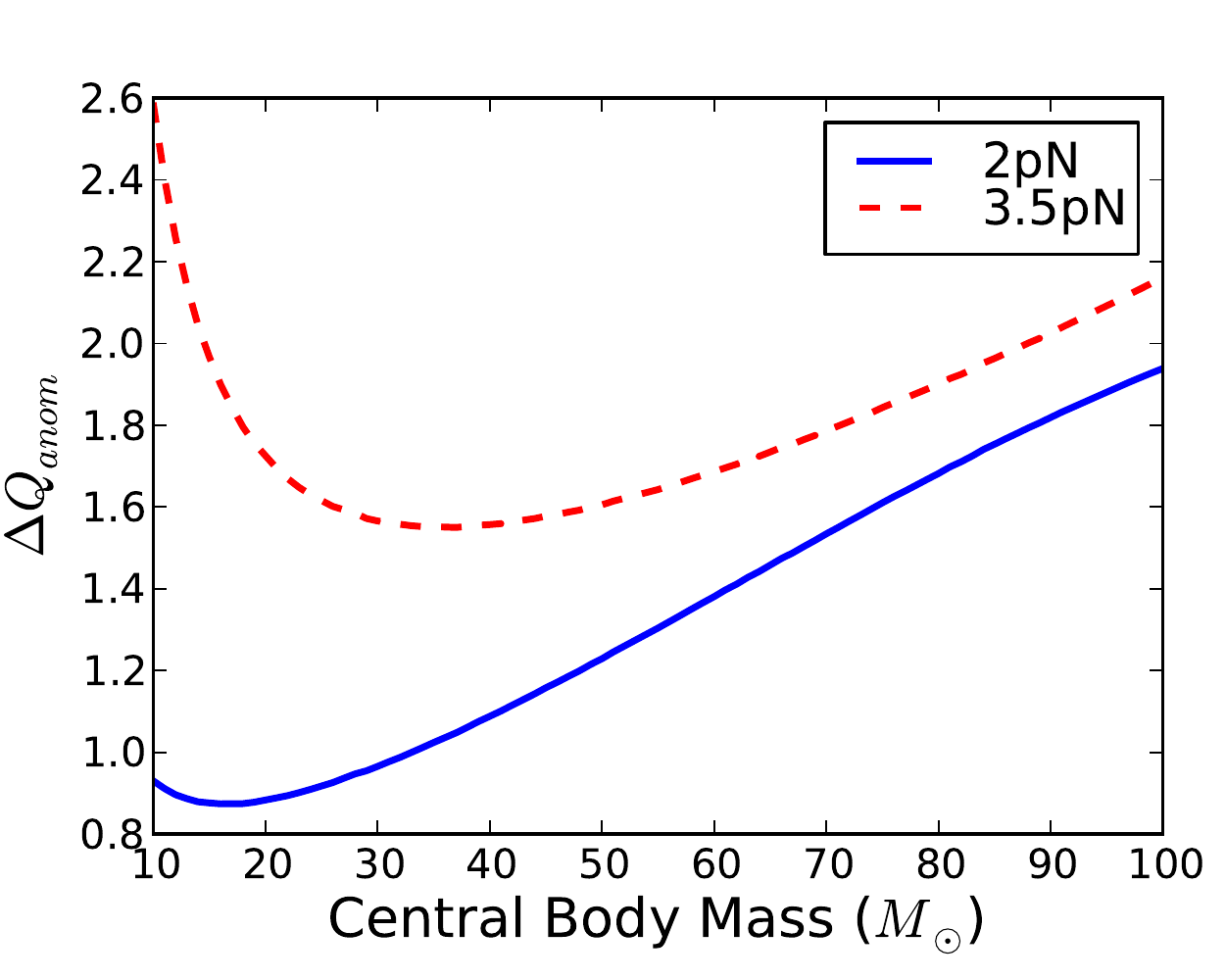}
  \caption{Standard deviations of $Q_{\text{anom}}$ as a function of central body mass (with $\chi_1 = 0.9$) for an IMBH-NS system (companion mass of $1.4M_{\odot}$).  The minimum standard deviation is waveform dependent (at $17M_{\odot}$ for 2 pN and at $37M_{\odot}$ for 3.5 pN).  In our dimensionless units, the quadrupole moment for these systems is $|Q|=0.81$.}
 \label{MvsQ}
\end{figure}

As a pure test of the no-hair theorem, we examine the accuracy of measuring the intrinsic parameters for the case when the best-fit $Q_{\text{anom}}=0$.  The results for our various IMRI systems are reported in Table \ref{nulltable}.  We study IMRIs with central body masses of 10, 25, 50, 100, and 150 $M_{\odot}$ \jontext{}{and central body spins of $0$, $0.5$ and $0.9$}, giving us a good look at the measurement accuracy as a function of both mass and spin.  It should be noted that although previous work looked at IMBHs up to 350 $M_{\odot}$, we found that with pN waveforms, the Fisher matrices were highly unstable numerically for these high-mass systems, with condition numbers well beyond the $10^{15}$ cutoff.  The detectability of waveform parameters improves the longer the IMRI signal remains in the detector band.  Thus, we find that higher $\chi_1$ typically leads to a more accurate measurement of $Q_{\text{anom}}$.  The third section ($\chi_1 = 0.9$) in Table \ref{nulltable} suggests that the ideal system lies somewhere in the $10M_{\odot}$ to $50M_{\odot}$ range, depending on the pN order of the waveform model.  In Fig. \ref{MvsQ}, we plot the standard deviations of $Q_{\text{anom}}$ as a function of central body mass for the highest spin case.

We find, as expected, that there does exist an ideal \jontext{massed}{mass of the} system for conducting a null test, depending on the order of the pN phase used.  In fact, our best systems for detecting $Q_{\text{anom}}$ are \jontext{medium massed}{} systems \jontext{}{of medium mass}, where the final \jontext{moments}{stages} of inspiral lie in the high sensitivity section of the Advanced LIGO bandwidth.  Since we have fixed the mass of the companion object in Fig. \ref{MvsQ} to 1.4$M_{\odot}$, changing the central body mass simultaneously adjusts the total mass and the mass ratio, making the observed minima in $\Delta Q_{\text{anom}}$ a competition between the ISCO frequency and the mass ratio.

\begin{figure}[h!]
\centering
\includegraphics[scale=0.7]{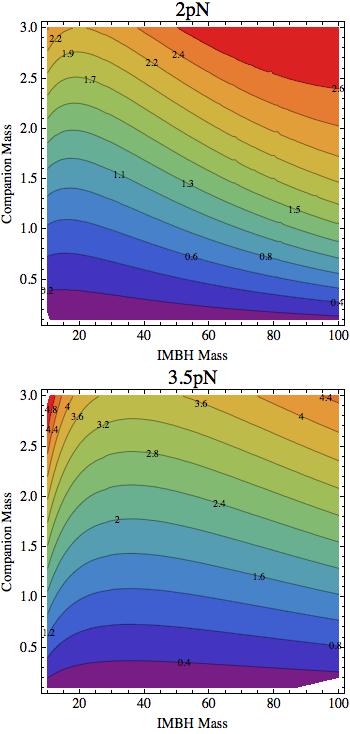}
\caption{$1$-$\sigma$ uncertainty estimates for $Q_{\text{anom}}$ as a function of  central body mass and the companion object mass. The lines indicate contours of constant $1$-$\sigma$ error in $Q_{\text{anom}}$. The structure of the plots is an interplay between the proximity of the ISCO frequency to detector's peak sensitivity (as determined by the total mass) and the preference for more extreme mass ratios to better probe the strong-field of the central body.  Both plots are for spin $\chi_1 = 0.9$; plots for other spin values are qualitatively similar, but decreasing the spin leads to an overall deterioration of accuracy and some shifting of the contour lines, due in part to the change in ISCO frequency.}
\label{MvsMvsQ}
\end{figure}

To reiterate, as the total mass of the system increases, our upper frequency cutoff (from the system ISCO, equation (\ref{ISCO})) decreases.  This moves the end of the waveform towards the most sensitive section of our advanced detector noise curve (250Hz).  With the final part of the waveform in the most sensitive part of the detector bandwidth, we become better able to discern physical effects that arise at higher pN orders, despite the overall loss of signal information (owing to the \jontext{}{reduction in the} time \jontext{spent}{the signal spends} in the detector \jontext{}{band} \jontext{,}{--- we adjust the distance so that} the SNR is kept fixed \jontext{}{as the mass is changed}).  At the same time, a more extreme mass ratio allows us to better ignore the contributions of the companion: the more asymmetric the ratio, the better the test particle approximation becomes, giving us a nearly geodesic map of the central body spacetime.  To better see this, we allow the mass of the companion object to vary, changing the total mass and mass ratio independently.  We plot the measurability of $Q_{\text{anom}}$ as a function of both masses in Fig. \ref{MvsMvsQ}.  Overall, the more asymmetric the mass ratio (towards the bottom right of the contour plots), the more accurately we can measure the mass-quadrupole moment; however, this only holds up to a certain total mass, and as the total mass pushes $f_{\text{ISCO}}$ below Advanced LIGO's peak sensitivity, the uncertainty on $Q_{\text{anom}}$ starts to increase again, as fewer cycles fall into the sensitive frequency band.

Given the large standard deviations, it will be difficult to precisely constrain the mass-quadrupole moment as an independent parameter.  Even \jontext{a system most sympathetic}{the systems best suited} to our purposes would only \jontext{permit}{provide} a 1-sigma \jontext{confidence}{error in $Q_{\text{anom}}$} on the order of the Kerr quadrupole moment.  This will limit the precision of tests \jontext{using}{of} the no-hair theorem, and we must await the detection of extreme-mass-ratio inspirals by a LISA type detector~\cite{BCEMRI} or the detection of IMRIs by a third-generation \jontext{}{ground-based} detector like the Einstein Telescope~\cite{HuertaIMRIA,HuertaIMRIB} for more precise tests.  Despite this difficulty, the detection of an IMBH with a best-fit value of $Q_{\text{anom}}\approx 0$ will allow us to rule out strong deviations from the Kerr geometry, such as those expected for boson stars.

\subsection{Off-Kerr Detectability}

\begin{figure*}[thb]
 \centering
 \includegraphics[scale=0.65]{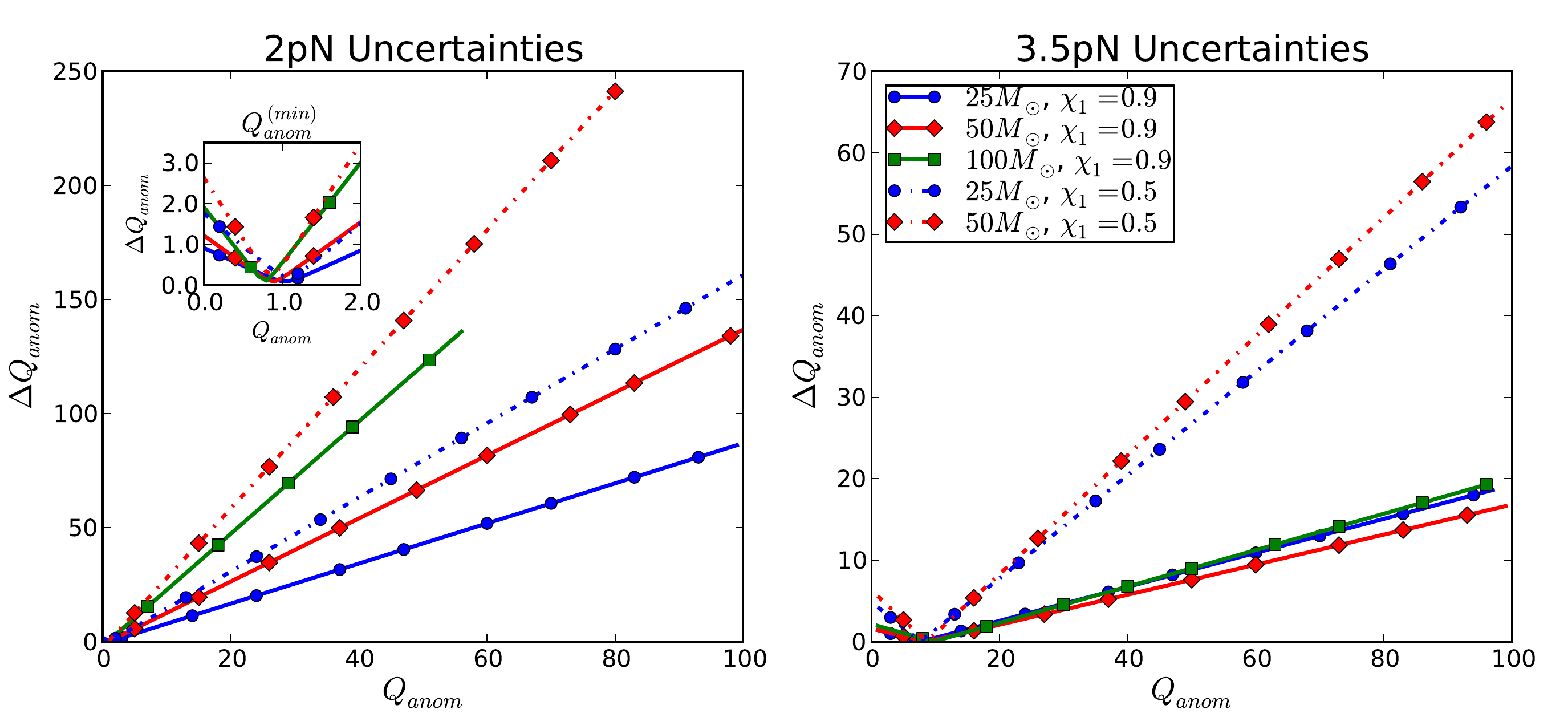}
  \caption{The $1$--$\sigma$ uncertainty in $Q_{\text{anom}}$ as a function of the best-fit value of $Q_{\text{anom}}$, for several choices of the central body mass and spin magnitude.  The solid lines are spin $\chi_1 = 0.9$ systems, and the dashed lines are $\chi_1=0.5$.  In both cases, the standard deviations dip to a local minimum, $Q_{\text{anom}}^{\text{(min)}}$, which varies with pN order.  Using the 3.5pN waveforms, the fractional error plateaus are sufficiently low that it should be possible to distinguish between a Kerr black hole and a more exotic object (e.g. a boson star).  The plots cut off the higher mass systems in the 2pN case and ignore the medium spin 100$M_{\odot}$ system, as these fall outside our numerical cutoff for stable inversion of the Fisher matrix.}
 \label{QvsSQ}
\end{figure*}

We now explore the potential to detect a quadrupole moment with a non-zero best-fit value of $Q_{\text{anom}}$.  In other words, \jontext{upon detection of}{if we detect} a signal suggesting a violation of the null-hypothesis that the central body is a Kerr black hole, how confident can we be of this violation?  We focus on lower mass systems (25$M_{\odot}$, 50$M_{\odot}$, and 100$M_{\odot}$ central bodies), with spins of  $0.5$ and $0.9$.  The results are plotted in Fig. \ref{QvsSQ}.  Again, as expected, the precision with which we can measure $Q_{\text{anom}}$ improves when the mass and spin values place the late stages of the inspiral into the most sensitive portion of the detector bandwidth.  Additionally, we find that the determination of all other parameters is, to lowest order, independent of the value of $Q_{\text{anom}}$.

For both waveform models, the fractional uncertainty in $Q_{\text{anom}}$ reaches a local minimum, at $Q_{\text{anom}}^{(\text{min})} \sim0.8$ for the 2pN waveform, and at $Q_{\text{anom}}^{(\text{min})}\sim10$ for the 3.5pN waveform.  The rise on either side of the dip is clearly a linear effect, with the asymptotic fractional errors $\Delta Q_{\text{anom}} / Q_{\text{anom}}$ being sufficiently small in some cases ($\sim15\%$ for highly spinning systems in the 3.5pN case) that large values of $Q_{\text{anom}}$ could easily be measured.  In fact, the measurement accuracy is much higher in the vicinity of $Q_{\text{anom}}^{(\text{min})}$.  At first, this effect seems \jontext{absurd}{somewhat surprising}: since $Q_{\text{anom}}$ is merely a phase factor, one might expect that different \jontext{best-fit}{} values would cause a minimal effect on parameter estimation.  %The fact that the uncertainty achieves a local minimum naively suggests some sort of failure in the Fisher matrix formalism.  

The explanation comes from an examination of the parameter correlations.  At the minimum uncertainty, the correlations between the two mass quantities and the anomalous quadrupole moment also tend towards zero.  As $Q_{\text{anom}}$ passes through the minimum, the signs of the correlations reverse, suggesting that at the dip, the projections of the multidimensional error ellipse onto our given choice of parameters are roughly spherical, since the correlation between parameters determines the eccentricity and orientation of the ellipses in error space. 

\begin{figure}[htb]
\centering
\includegraphics[ scale=0.8]{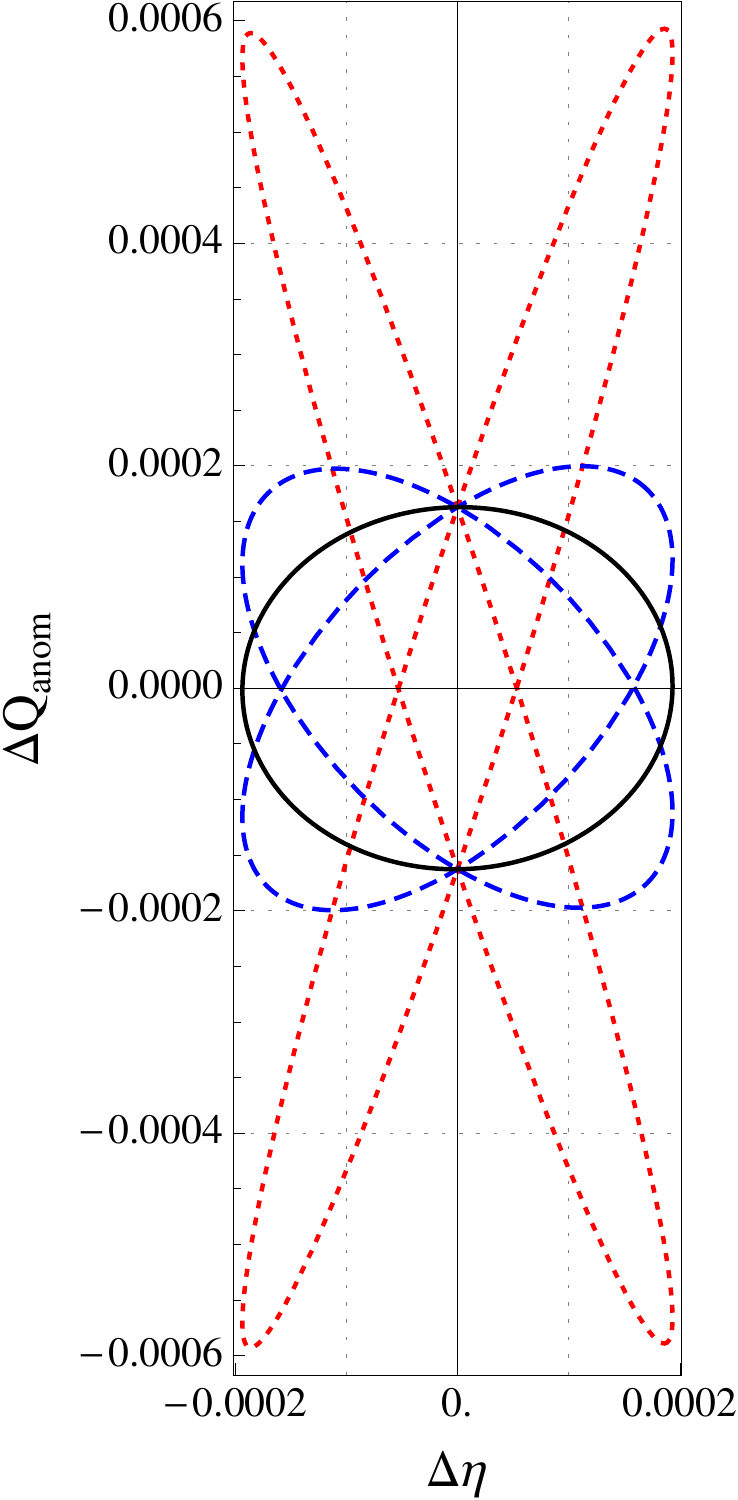}
\caption{2-dimensional error ellipses from a $\eta$-$Q_{\text{anom}}$ Fisher matrix (with other parameters held fixed), evaluated near the value of $Q_{\text{anom}}=Q_{\text{anom}}^{(\text{min})}$ (in solid black).  The colored ellipses are small deviations from the minimum ($\pm 0.1$ in dashed blue, and $\pm 0.5$ in dotted red).}
\label{ellipses}
\end{figure}

This is confirmed by looking at the volume of the $\mathcal{M}_c$-$\eta$-$Q_{\text{anom}}$ sub-matrix of the full correlation matrix.  As expected, the volume of the error ellipsoid is conserved, independent of the \jontext{best-fit}{} value of $Q_{\text{anom}}$.  To demonstrate the eccentricity/orientation effect, we examine a 2-dimensional Fisher matrix, using only $\eta$ and $Q_{\text{anom}}$ as our parameters, and use the inverse of the Fisher matrix, $\Sigma$, to plot the error ellipses in the $\eta$-$Q_{\text{anom}}$ coordinate system.  A representative number of these ellipses are shown in Fig. \ref{ellipses}.

As the value of $Q_{\text{anom}}$ is varied away from $Q_{\text{anom}}^{(\text{min})}$, we see that the error ellipses rapidly deform, going from nearly circular at the minimum to highly eccentric and rotated.  As the quadrupole moment is further changed from the minimum, the projection of the error ellipse onto the  $Q_{\text{anom}}$ axis increases, while the projection onto the $\eta$ axis remains the same.  Although the eccentricity and orientation of the ellipses vary, the area of the ellipses is constant.  This is to be expected, since $\sigma_{qm}$ depends on a combination of the masses and the quadrupole moment.   Furthermore, an analytic examination of the Fisher matrix elements shows that only the $\mathcal{M}_c$ and $\eta$ rows depend on the injected values of  $Q_{\text{anom}}$.  We conclude that the orbital evolution is not directly affected by the quadrupole moment; only via correlation with the masses does the value of $Q_{\text{anom}}$ affect parameter estimation. 

This deformation of the mass/quadrupole moment ellipsoids suggests that the correlation between the two mass parameters and the anomalous quadrupole moment is of greatest importance for parameter estimation of $Q_{\text{anom}}$.  This effect is also responsible for the increase in the measurement accuracy of large $Q_{\text{anom}}$ at higher pN orders:  since the 3.5 pN waveforms allow for more accurate determination of system masses \cite{ArunPE}, they also decrease the volume of the error ellipsoids in our $\mathcal{M}_c$-$\eta$-$Q_{\text{anom}}$ space, thereby decreasing the possible projection of the standard deviations onto the $Q_{\text{anom}}$ axis.  This effect was confirmed by a direct comparison of the error volume between the 2pN and 3.5pN waveforms.  Figure \ref{QvsSQ} can be seen as one-dimensional projections of this effect. 

%\jon{OK, so the argument here is that as $Q_{\text{anom}}$ changes, the correlation between that parameter and the chirp mass and mass ratio change, but the precision with which the combination can be measured is unchanged. Why does the correlation change? Does the minimum point correspond to $\sigma_{qm}$ becoming comparable in size to the other terms in the expression for $\alpha_4$ (in the 2pN case which is a bit easier to understand)? In some of the papers by Sathya et al., they look at the precision of measuring the various PN phase coefficients and discuss their correlations, so maybe those results could reinforce what you are saying. It may be that the correlation between $\alpha_4$ and the other $\alpha_i$'s depends on the value of $\alpha_4$ and that is what you are seeing here. I'm not suggesting lots of extra work, but looking at those papers and adding a reference to them could be useful.} \carl{Sathya's FIM paper on various pN orders was actually the paper I used to check our code, so I'm pretty familiar with it.  I was arguing that the improvement in mass measurement and 3.5 vs 2 pN decreases the overall volume of our ellipses (I changed the wording in the above paragraph to reflect that better, and cited the paper).  A quick glance at the Arun et al shows that their mass correlations only change from 0.95 to 0.93 from 2 to 3.5 pN, so I'm unsure that the mass correlations by themselves are a significant effect.}

Given this understanding of the parameter correlations, we believe that it will be possible, under favorable conditions, to detect a large off-Kerr quadrupole moment in IMRI signals detected by Advanced LIGO, especially if the best-fit value of $Q_{\text{anom}}$ happens to lie near $Q_{\text{anom}}^{\text{(min)}}$.    This explains the increased effectiveness of the 2pN waveforms at performing the null test that was noted earlier: since $Q_{\text{anom}}^{\text{(min)}} \approx 0.8$ for the 2pN waveforms, the overall error ellipsoid at $Q_{\text{anom}} = 0$ is closer to spherical than in the 3.5pN case, yielding a smaller projection onto the $Q_{\text{anom}}$ axis, even though the total error volume is greater.  However, a precision measurement will require a system with a highly spinning central body and a relatively small central mass.  In order to reliably test for exotic stars in Advanced LIGO, we will, above all, require accurate IMRI waveforms.

\subsection{Sources of Systematic Errors}

There are a number of technical issues that can arise \jontext{given}{in} our analysis.  We now discuss both the disadvantages of our formalism as well as the problems inherent in the waveform model.  We have already touched on the generic issues with the Fisher matrix formalism in section IIIA; what we discuss now are the specific issues that arise from parameter correlations.  We also consider the suspect nature of introducing large perturbations into pN terms (i.e., $Q_{\text{anom}}$), as well as the dependence of our results on assumptions regarding the low-frequency sensitivity of advanced detectors.

First, recall from our introduction to the Fisher matrix the issues that arise from high parameter correlations: when two or more parameters become highly correlated, the submatrix between those parameters tends towards singular.  As we move \jontext{towards more}{into} certain areas of our parameter space (in our case, higher total masses), we find that the correlations and condition numbers increase\jontext{ with total mass}{}.  An example of this effect is \jontext{detailed}{shown} in Table \ref{cctable}.  This demonstrates that our inversion criterion for the Fisher matrices was chosen well, allowing us to avoid rounding errors from inverting matrices beyond \jontext{the}{} machine precision.  The fact that most of our systems yield high parameter correlations suggests that the FIM must be used with care in this mass region of the parameter space. 

\begin{table}[bthp]
\caption{The mean and maximum cross-correlation coefficients, $C_{ij}$, in the Fisher matrix for prototypical systems (with $m_2 = 1.4M_{\odot}$ and 3.5pN in phase).  While the maximum $C_{ij}$ decreases with an increase in total mass (corresponding to a decrease in $1-{\rm Max}(|C_{ij}|)$), the average of all correlations increases with total mass.  The one exception is the non-spinning $m_1 = 150M_{\odot}$ case, but this had a covariance matrix with a condition number roughly 5 times larger than our numerical cutoff.  All other results are roughly at or below our inversion tolerance.  As the mean correlation increases towards one, the matrices become more susceptible to numerical rounding errors, indicating the FIM is ill-suited in the high mass region of parameter space.}
%\jon{I think it is OK to include the table, but you need to explain what it shows better.}\carl{ok, I've cleared up the text here}}
\begin{tabular}{lcccc}

\hline\hline
Central Body & \vline &  Mean$(|C_{ij}|)$ & \vline & $1-$Max$(|C_{ij}|)$\\
\hline
$\chi_1 = 0$\\
\hline
$10M_{\odot}$ & \vline & 0.92  &\vline & $6.7\times10^{-6}$\\
$25M_{\odot}$ & \vline & \jontext{0.69}{0.96} &\vline & $3.1\times10^{-5}$\\
$50M_{\odot}$ & \vline & 0.97 &\vline & $4.4\times10^{-5}$\\
$100M_{\odot}$ & \vline & 0.98 &\vline & $8.0\times10^{-5}$\\
$150M_{\odot}$ & \vline & 0.92 &\vline & $3.5\times10^{-4}$\\
\hline\hline
$\chi_1=0.9$\\
\hline
$10M_{\odot}$ & \vline & 0.79 &\vline & $1.3\times10^{-4}$\\
$25M_{\odot}$ & \vline & 0.93 &\vline & $2.7\times10^{-4}$\\
$50M_{\odot}$ & \vline & 0.97&\vline & $3.8\times10^{-4}$\\
$100M_{\odot}$ & \vline & 0.98 &\vline & $4.6\times10^{-4}$\\
$150M_{\odot}$ & \vline & 0.99 &\vline & $5.2\times10^{-4}$\\

\hline\hline

\end{tabular}
\label{cctable}
\end{table}

In addition to the issues with the Fisher matrix, the post-Newtonian waveforms also \jontext{function as a source of}{give rise to} systematic errors in our study.  We (Mandel and Gair, \cite{MandelGairDetection}) have previously discussed the problems of convergence and breakdown in the pN waveforms as applied to IMRIs.  The problem \jontext{we concern ourselves with now}{in the present case} arises at large values of $Q_{\text{anom}}$, since increasing the pre-factor of the 2pN term increases the contributions to the orbital evolution from higher order terms.  In some cases (e.g. $Q_{\text{anom}} \sim 100$), the 2pN effects from the quadrupole moment are significant relative to the lowest order Newtonian orbital effects. This problem can be quantified by measuring the number of orbital cycles each term contributes to the inspiral.  To that end, we look at the phase in the time domain up to 2pN order,

\begin{equation}
\Phi(f) = \phi_c - \frac{1}{16 \eta}v^{-5} \sum^{4}_{i=0}a_i v^i
\end{equation}

\noindent where $f$ is the gravitational wave frequency (see Poisson and Will, c.f. equation 3.2), and $v=v(f) = (\pi M f)^{1/3}$ as before.  The coefficients are

\begin{align}
a_0 &=  1 \\
a_1 &= 0 \nonumber\\
a_2 &=  \frac{5}{3}\left(\frac{743}{336} + \frac{11}{4}\eta\right)\nonumber\\
a_3 &= - \frac{5}{2}(4\pi - \beta)\nonumber\\
a_4 &= 5\left(\frac{3058673}{1016064} + \frac{5429}{1008}\eta + \frac{617}{144}\eta^2 -  \sigma_{qm}\right)\nonumber
\end{align}

\begin{figure}[tbh]
 \centering
 \includegraphics[scale=0.69]{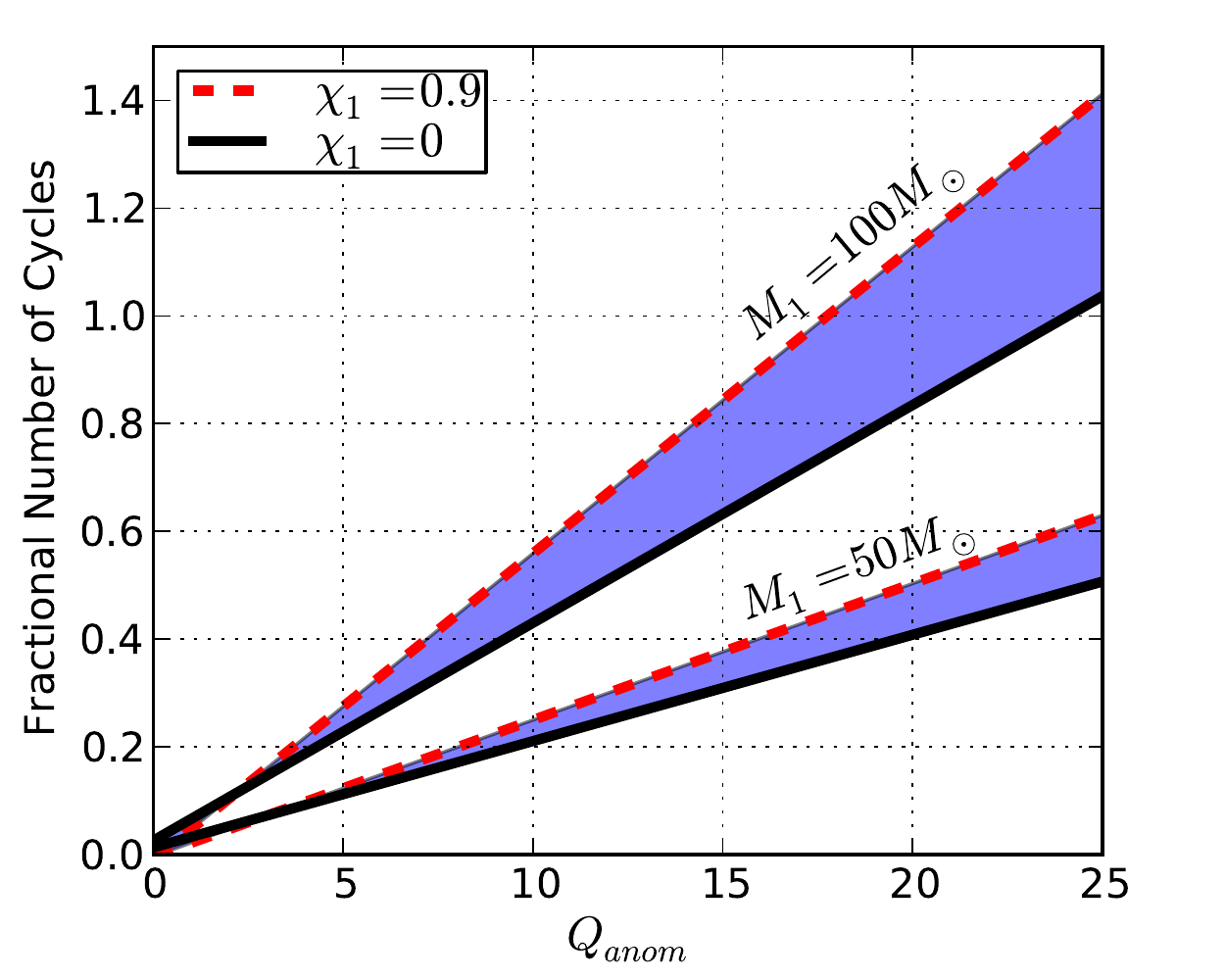}
  \caption{The fraction of orbital cycles contributed by the 2pN verses the 0pN terms as a function of the anomalous quadrupole moment.  The two shaded regions are for two typical IMRI systems with spins ranging from 0 (black) to 0.9 (red).  Note that the initial cross over between the non-spinning and highly spinning cases arises from the sign change in $\alpha_4$ at low $Q_{\text{anom}}$. The validity of the approximation that ignores higher-order $Q$-dependent terms comes into doubt when the number of fractional cycles is high.} 
 \label{fraccycles}
\end{figure}

We then compare the fraction of cycles generated by the $0^{\rm th}$ (Newtonian) and the $2^{\rm nd}$ order term from the frequency the signal enters the detector ($f_0 = 10 \text{Hz}$) to the ISCO frequency of the system ($f_{\text{ISCO}}$).  Explicitly,

\begin{equation}
R = \frac{a_4}{a_0}\left[ \frac{ v^{-1}(f_{\text{ISCO}}) - v^{-1}(f_0)}{v^{-5}(f_{\text{ISCO}}) - v^{-5}(f_0)}\right ]
\label{numbercycles}
\end{equation}

\noindent gives the fraction of cycles.  The pre-factor, $a_4/a_0$, depends on the physical parameters of the system, so we can look at how close to unity equation~(\ref{numbercycles}) can become for various systems.  \jontext{See}{This is shown in} Fig. \ref{fraccycles}.  It is clear from this plot that the validity of the pN expansion given our definition is inversely proportional to the value of $Q_{\text{anom}}$.  This is in addition to the problem of using pN waveforms for such an extreme mass ratio.  Fig. \ref{fraccycles} \jontext{confirms}{indicates} that even low mass systems can be pushed \jontext{to invalid ranges}{beyond their range of validity} with a sufficiently large physical value of $Q_{\text{anom}}$.

One could make the argument that having a greater number of cycles generated at 2pN than at 0pN is not necessarily invalid; after all, since we are including both terms, such a large fraction would simply represent a regime in which Newtonian orbital effects are \jontext{dominated by}{subdominant to} quadrupole-coupling effects.  However, we have only included the effects of the quadrupole moment to lowest known order.  If the 2pN term dominates at high values of  $Q_{\text{anom}}$, there is no reason \jontext{}{to suppose} that the unknown contributions at higher orders will not be similarly dominant in the ``true'' waveform.  We once again conclude that any precision tests, particularly of exotic objects with a large quadrupole moment, will require new waveforms (most likely from perturbation theory and/or effective-one-body approaches) that have been specifically tuned for the IMRI regime.

\begin{table}[bthp]
\caption{As in the bottom right section of Table \ref{nulltable}, this table shows standard deviations on $Q_{\text{anom}}$ for $m_2 = 1.4$, and using the 3.5pN in phase waveform model.  We change the lower bound on frequency integration to explore how the results scale with the Advanced LIGO low-frequency sensitivity cutoff.  The SNR is kept fixed at $\rho=20$\jontext{, so that the total information is kept constant}{}.}
\begin{tabular}{lcccccc}

\hline\hline

$\chi_1 = 0.9$ & \vline & \multicolumn{5}{c}{$\Delta Q_{\text{anom}}$} \\
\hline
Central Body & \vline &  $f_0 = 10$Hz & \vline & $f_0 = 20$Hz & \vline & $f_0 = 40$Hz\\
\hline
$10M_{\odot}$ & \vline & 2.59 &\vline & 3.93 & \vline & 7.89 \\
$25M_{\odot}$ & \vline & 1.62 &\vline & 3.05& \vline & 7.57\\
$50M_{\odot}$ & \vline & 1.61 &\vline & 3.65& \vline & 6.67\\
$100M_{\odot}$ & \vline & 2.16 &\vline & 4.74& \vline & 6.95\\

\hline\hline

\end{tabular}
\label{frequencytable}
\end{table}

Finally, a large part of the information extracted from our IMRI signals comes from low frequencies, and therefore depends on assumptions regarding the low-frequency sensitivity of the detectors.  While we have used the current best guess for the power spectral density\jontext{ available}, it is quite possible that Advanced LIGO will not achieve the lower frequency sensitivity that has been targeted.  In that case, significant amounts of information will be lost from the shift in the lower frequency cutoff, as many in-band orbital cycles would become inaccessible to observation.  We can look at the decreased sensitivity to $Q_{\text{anom}}$ that arises from a different lower frequency bound in the signal inner product (equation (\ref{overlap})).  The results, in Table \ref{frequencytable}, confirm our statement that the information from lower frequencies is key to measuring higher order pN effects.  The low-frequency sensitivity of Advanced LIGO will be vital for the tests discussed in this paper to be performed.

\section{Conclusion}

We summarize here the key results of this paper.  Using frequency domain post-Newtonian waveforms, we tested the possibility of measuring the quadrupole moment of a massive, spinning compact body as a parameter independent of the object's mass and spin.  This served two distinct purposes:  first, a simultaneous measurement of the mass, spin, and quadrupole moment of a Kerr black hole would serve as a test of the no-hair theorem of general relativity, acting as a check on the internal consistency of the theory.  Secondly, the detection of an off-Kerr quadrupole moment would suggest the existence of some new, exotic type of star or other supermassive object, allowing Advanced LIGO to identify possible new forms of matter in the Universe. 

For this analysis, we first explored whether we would be able to rule out significant deviations from a Kerr black hole in the case when the best-fit parameters correspond to $Q_{\text{anom}} = 0$.
For this null test, we found that the ability of Advanced LIGO to measure an anomalous quadrupole moment will depend heavily on the parameters of the system.  The ideal systems are medium mass, highly spinning systems with an ISCO frequency close to $250$Hz, for which the final moments of the inspiral signal lie at frequencies in the highest-sensitivity range of the detector.  Additionally, since systems with more extreme mass ratios spend more orbital cycles in the strong-field of the central body, their inspirals will encode more information about the quadrupole moment of the spacetime.  It is the interplay between these two effects, the ISCO frequency and the mass ratio, that determines the best system for these tests.  The optimal system also depends on the pN order of the waveform used, due to the increased mass precision at higher orders and the correlation between the masses and the quadrupole moment of the central body.

If the central body has a non-zero quadrupole moment, the fractional uncertainty with which $Q_{\text{anom}}$ can be measured varies as the $\mathcal{M}_c$-$\eta$-$Q_{\text{anom}}$ error ellipse is slowly deformed.  The error has a local minimum at $Q_{\text{anom}}^{\text{(min)}}$, representing an error ellipse who principal axes nearly align with the $\mathcal{M}_c$-$\eta$-$Q_{\text{anom}}$ coordinates, and increases linearly with $|Q_{\text{anom}}-Q_{\text{anom}}^{\text{(min)}}|$.  The slope of this linear effect is sufficiently small that certain systems could be easily identified to within 15\% accuracy for large values of $Q_{\text{anom}}$.  If one of these near-optimal systems was observed, Advanced LIGO would be able to definitively rule out that it was a Kerr black hole.

\jon{The space-based gravitational wave detector (e)LISA will also have the capability to measure anomalous quadrupole moments of black holes. Barack and Cutler~\cite{BCEMRI} showed that a LISA observation of an EMRI with a signal-to-noise ratio of $100$ would be able to detect a deviation in the quadrupole moment of $10^{-4}$--$10^{-2}$, depending on the system parameters, when the central object is a Kerr black hole. If the effect we described in this paper, that the quadrupole moment can be measured more accurately when it is not equal to the Kerr value, also holds for EMRI systems, (e)LISA could potentially do even better. However, that analysis needs to be revisited in light of the recent redesign of the eLISA mission~\cite{LISAdescope}, which has led to a change in the frequency sensitivity of the mission and has meant that an SNR of $100$ is now rather optimistic for EMRI sources. eLISA has the potential to provide more precise constraints than Advanced LIGO because EMRI systems will be observed for many more cycles in the strong-field regime than IMRI systems. However, even optimistically, eLISA will not be producing data until $5$--$10$ years after Advanced LIGO. The results in this paper indicate that Advanced LIGO could provide the first hints for departures from the Kerr metric, which would then be confirmed with more precision by future detectors including eLISA.}

The primary limitations in the study presented in this paper were the lack of accurate  waveforms covering the IMRI mass regime and the limitations of the FIM approach.  In order to properly study the parameter space, we would require a realistic signal from an IMRI template, studied with a \jontext{more Bayesian complete}{Bayesian} technique (e.g.\ Markov Chain Monte Carlo).  The primary limitation on our ability to measure $Q_{\text{anom}}$ is the correlation in the mass/quadrupole subspace. We would therefore expect to measure $Q_{\text{anom}}$ more precisely if we were able to measure the masses more precisely. We have used a pN approximant waveform in this study, so we may be missing pieces of the frequency evolution that would help break this degeneracy. A further study using a more realistic waveform model would be required to investigate if we could do better in practice than the results described here. \carl{Current work is underway in developing such waveforms via the ``effective one-body -- numerical relativity'' technique (see \cite{PanEOBNR}), though these studies have yet to produce fully spinning IMRI waveforms in the strong-field regime.}  However, even using this simplified waveform model, we find that we would be able to differentiate between black holes, boson stars, and naked singularities in the Advanced LIGO era, which would represent a significant achievement for compact-object physics.  \carl{Assuming IMRI waveforms will be available in the next few years, the realistic measurements may be even more precise than the results quoted here.}

Future work will be needed to quantify the biases introduced by our use of a pN approximant. While we have explored some of these  errors by comparing different pN orders, there also exists the bias introduced by modeling a waveform with $Q_{\text{anom}}$ present when it does not exist in nature (or the converse).   This work can also be extended to proposed third generation detectors (e.g.\ the Einstein telescope \cite{JonET}), which could offer increased sensitivity at the same SNR due to their lower low-frequency cutoff.

\section*{Acknowledgements}

The authors would like to thank Duncan Brown, Vicky Kalogera, Gregory Oliver and Alberto Vecchio for useful discussions.  CR was partially supported by the Illinois Space Grant Consortium Fellowship and by an NSF GRFP Fellowship, award DGE-0824162 .  IM was partially supported by the NSF Astronomy and Astrophysics Postdoctoral Fellowship, award AST-0901985. JG's work is supported by the Royal Society.

\bibliography{qanom_paper}{}

\end{document}